\documentclass{article}
\title{Project Report: Achieving Efficient and Provably Secure Steganography in Practice}
\author{Aubrey Alston (ada2145@columbia.edu)}

\usepackage[left=2.5cm,top=3cm,right=2.5cm,bottom=3cm,bindingoffset=0.5cm]{geometry}

\usepackage{fancyhdr}
\date{}

\usepackage{amsmath}
\usepackage{amssymb}
\usepackage{MnSymbol}
\usepackage{graphicx}
\usepackage{float}

\usepackage{algorithm}
\usepackage{algpseudocode}
\usepackage{algorithmicx}
\usepackage{program}

\begin{document}

\maketitle

\section{Overview}

Steganography is the task of concealing a message within a medium such that the presence of the 
hidden message cannot be detected.  Though the prospect of steganography is conceivably interesting 
in many contexts, and though work has been done both towards formalizing steganographic security and 
providing provably secure constructions, little work exists attempting to provide efficient and provably 
secure steganographic schemes in specific, useful domains.  

Beginning from the starting point of the initial definition of steganographic security, I have 
engaged in an exploration which has developed to include two primary tasks, both pointing towards 
the realization of efficient and secure steganographic systems in practice: (a) investigating the 
syntactic and semantic applicability of the current formalism of steganographic security 
to a broader range of potentially interesting domains and (b) constructing and implementing provably 
secure (symmetric-key) steganographic schemes in domains which are well-suited to the current formalism.
\newline\newline
\noindent In the remainder of this document, I provide a high-level overview of existing work in the 
area of provably secure steganography, and I describe the progress I have made in the tasks stated above.
\footnote{This work is the report authored in conjunction with research performed with the Columbia University
cryptography laboratory.}

\tableofcontents

\section{Paper: Understanding Provably Secure Steganography}

In 2002, Hopper et. al. published ``Provably Secure Steganography,'' a work studying steganography 
from a complexity-theoretic perspective which set forth the current formal definition of provable 
steganographic security \cite{BiglouPSS}.  As such, the content of this work is vital to any 
exploration seeking the realization of provably secure steganography in practice.  What follows in this 
section is a summary of this work, including the formal definition of steganographic security which it defines.
\newline\newline
In this work, the authors formulate steganography as a game involving three parties: Alice, Bob, and Ward.
In this game, Alice attempts to pass a secret message to Bob, and Ward is attempting to determine whether 
or not Alice is sending a secret message.

Towards formally defining security in this setting, the authors abstract communication between Alice and 
Bob as occurring over a \textbf{channel}.  A \textit{channel} $\mathcal{C}$, formally, is a statistical distribution on bit sequences where each bit is marked with monotonically increasing time values: a statistical distribution with 
support $(\{0,1\}, t_1) ... (\{0,1\}, t_n)$, $\forall i, t_{i+1} \geq t_i$ \footnote{This definition 
of a channel is chosen specifically by the authors so as to be able to incorporate the notion of time; this 
allows for timing attacks to be included within the formal definition of security.}.

Communication over a channel is an action which requires one or more parties to draw from a channel.  Specifically, 
the authors assume the existence of an oracle capable of drawing fixed-length bit sequences from the channel conditioned on a history $h$ of bits already drawn.  The authors denote by $\mathcal{C}_h$ the channel distribution conditioned on history $h$, and the authors denote by $\mathcal{C}_h^{b}$ the conditional distribution over the next $b$ bits (a block) drawn from the channel, again conditioned on $h$.  For any steganographic construction under this formalism, the authors 
state that $b$ must be fixed, and the authors require that, for all potential histories $h$, the minimum entropy 
of $\mathcal{C}_h^{b}$ be greater than 1.\footnote{Equivalently stated, the block $B$ with maximum probability over the 
support of $\mathcal{C}_h^{b}$ must have $Pr[B] < 0.5$}
\newline\newline
The authors define a \textbf{stegosystem} as a pair of probabilistic algorithms $S = (SE, SD)$.  $SE$ takes 
as input a key $K \in \{0,1\}^k$, a \textit{hiddentext} $m \in \{0,1\}^*$, a message history $h$, and an 
oracle $M(h)$ which samples blocks according to $\mathcal{C}_h^{b}$.  The result of $SE^{M(h)}(K, m, h)$ should 
be a sequence of blocks $c_1 \mid\mid c_2 \mid\mid ... \mid\mid c_l$, called the \textit{stegotext}, from the support 
of $C_h^{b*l}$.  $SD$ takes as input a key $K$, a stegotext $s$, a message history $h$, and returns as a 
result a hiddentext $m$.

At this point in the work, the authors discuss some of their choices in their formulation of the 
definition of a stegosystem.  Namely, the authors note that this formalism is constructed 
under the assumption that stegosystems do not take advantage of additional knowledge of the channel, resulting 
in (among other things) dependence on the use of the channel oracle.  The authors also note that 
this dependence may not be justified in all cases, as there are reasonable circumstances under which 
(a) the oracle draws are not independent or are not efficient (e.g. a human oracle) and/or (b) potentially 
$SD$ may not practically have access to an oracle.
\newline\newline
\noindent \textbf{Correctness of a stegosystem } The authors say that a stegosystem $S = (SE, SD)$ is correct if the following holds:

\[ \forall m, \mid m \mid < p(k): Pr[SD^M(K,SE^M(k,m,h),h) = m] \geq \frac{2}{3} \]

\noindent \textbf{Steganographic secrecy with passive wardens } The setting for 
steganographic secrecy against passive wardens is the  following game played by a 
warden $W$:

\begin{enumerate}
\item{$W$ has access to an oracle $M(h)$ which is able to sample blocks from the distribution 
$\mathcal{C}_h^{b}$ for arbitrary histories $h$.  $W$ is allowed to make as many draws from $M(h)$ as 
desired.}
\item{$W$ is given access to a second oracle which is either $SE(K,m,h)$ or a generic oracle 
$O(m,h)$ which samples from $\mathcal{C}_h^{\mid SE(K,m,h)\mid}$.  $W$ makes at most $q$ queries to this 
oracle, totaling $l$ bits of hiddentext.}
\item{$W$ outputs a bit which signals whether the output of the second oracle contains concealed 
messages.} 
\end{enumerate} 

In this setting, the advantage of an adversary (warden) against a stegosystem $S=(SE,SD)$ defined for a 
particular channel $\mathcal{C}$ is 

\[ Adv_{S,\mathcal{C}}(W) = \mid Pr_{K,r}[W_r^{M,SE(K,m,h)} = 1] - Pr_{r}[W_r^{M,O(m,h)} = 1] \mid \]

We would then say that a stegosystem is \textit{steganographically secret} against $(t(k), q(k), l(k))-$ wardens 
if the maximum advantage of any warden making $q(k)$ oracle queries totalling at most $l(k)$ bits of hiddentext and running in time at most $t(k)$ is negligible in $k$, where $k$ is the security parameter of the system.  Here, the authors attempt to define a notion of \textit{universal steganographic secrecy}: a stegosystem $S=(SE,SD)$ is \textit{universally stegnanographically secret} if, for every channel distribution $\mathcal{C}$ having minimum entropy greater than one and for every probabilistic polynomial time warden $W$, $Adv_{S_\mathcal{C}}(W)$ is negligible in $k$.
\newline\newline
\noindent \textbf{An example construction } Hopper et. al. provide multiple constructions which they prove to 
be secure in the passive warden model.  The first they provide is a stateful scheme which they prove to 
be universally steganographically secret conditioned upon the use of a pseudo-random function (PRF).

The scheme provided assumes that Alice and Bob both have access to a channel oracle $M(h)$ and a synchronizeed 
$d$-bit counter and that Alice and Bob have access to a PRF $F: \{0,1\}^k \times \{0,1\}^d \times (\{0,1\},t)^b \Rightarrow \{0,1\}$.  Further, the scheme assumes the existence of an error-correcting code able to correct a $\frac{3}{8} + negl(n)$ fraction of errors, the procedures of which are denoted by $Enc$ and $Dec$. The exact functionality of $SE$ and $SD$ are given below:

\begin{algorithm}[H]
\caption{Steganographic encoding procedure}\label{1a}
\begin{algorithmic}[1]
\Procedure{$SE$}{$K \in \{0,1\}^k, m', h$}
\State $m = Enc(m')$
\State Split $m$ into bits $m_1 \mid\mid m_2 \mid\mid ... m_l$
\For{$i$ from $1$ to $l$}
\State Sample $M(h)$ at most two times until $F_K(N,M(h))=m_i$.  Set $c_i = M(h)$ for the final sample.
\State Set $h = h\mid\mid c_i$.
\State Increment counter $N$.
\EndFor
\State \textbf{return} $c_1 \mid\mid...\mid\mid c_l$
\EndProcedure
\end{algorithmic}
\end{algorithm}

\begin{algorithm}[H]
\caption{Steganographic decoding procedure}\label{1a}
\begin{algorithmic}[1]
\Procedure{$SD$}{$K \in \{0,1\}^k, c, h$}
\State Split $c$ into $b-bit$ blocks $c_1 \mid\mid ... \mid\mid c_l$.
\For{$i$ from $1$ to $l$}
\State Set $m_i = F_K(N, c_i)$.
\State Increment $N$.
\EndFor
\State Set $m = m_i \mid\mid ... \mid\mid m_l$.
\State \textbf{return} $Dec(m)$
\EndProcedure
\end{algorithmic}
\end{algorithm}

Given the minimal entropy assumption of the channel, the authors are able to 
succinctly derive an upper bound on the failure probability (per message) bit 
of approximately $\frac{3}{8}$, allowing the scheme to be correct given the use 
of a satisfactory error correcting code.  The authors then show that the steganographic 
secrecy of this construction under the passive warden model directly reduces to the 
PRF security of $F$, thereby allowing them to conclude that this scheme $S$ is both correct 
and universally steganographically secret.
\newline\newline
\noindent \textbf{Other notions } The authors addtionally define additional formal 
notions for steganography, including the notion of \textit{robust steganography}.
This definition is an attempt to incorporate the idea of an active warden 
who modifies messages as they are sent; however, as the authors admit, an unbounded 
adversary may choose to simply replace messages completely from the channel distribution,
destroying concealed messages.  Their specific solution is to bound the warden by a 
relation of permissible modifications, but the practical relevance of these limitations 
is not directly clear.

\subsection{Initial Thoughts and Questions}

``Provably Secure Steganography,'' aside from presenting a useful formal framework 
for defining secure steganography, coerced from me an initial set of thoughts and 
questions surrounding the intersection of provably secure steganography and 
its realization in practice.  In this section, I summarize and discuss these thoughts
 and questions.

\subsubsection{The Notion of Universal Steganography}

The formal model presented by Hopper et. al. explicitly 
and implicitly portrays some desire to encapsulate as many potential channels and domains 
as possible.  On this topic, I am left with two questions; namely, (1) in attempting to achieve such generality, 
are some domains left out or made more difficult to manipulate within the framework?  and (2) are universal steganographic constructions under this model the most valuable contribution in practice?

Regarding the first of these questions, clearly the definition of `universal' disqualifies some channels, given 
the minimum entropy assumption enforced.   Additionally, one 
result of the generality of the formalism is the assumption that the specific characterization of the 
channel is unknown but that sampling is feasible: could there potentially be a useful channel 
where the characterization is known but sampling many times is not feasible?  The reliance on fixed block sizes 
also seems to add difficulty in domains where variable-length messages are common: a proof of security for a steganographic 
scheme within such a domain would require the added difficulty of proving minimum entropy while dividing 
messages along fixed-size block boundaries.

Regarding the second of these questions, consider the simple stateful universal steganographic 
scheme given in the original paper.  This scheme remains sufficiently general to apply to 
any channel which, under the correct choice of $b$, follows the minimum entropy assumption.  In using this scheme,
I may achieve a rate of secrecy of one bit per block.  Intuitively, however, it seems that there should exist an optimal scheme given an arbitrary channel $\mathcal{C}$, even for fixed $b$, that uses channel-specific properties to achieve a much better rate.  

To illustrate this intuition, say we model the output of Alice as a stochastic process which obeys the distribution $C_h^{b}$; when Alice is preparing to send symbol $i$, there exists some number $r_i \leq 2^b$ of potential symbols she may output while obeying the channel distribution.  In theory, then, it seems reasonable to believe that Alice should be able to hide $log_2(r_i)$, potentially greater than 1, bits of hiddentext in that symbol.  In the same manner, this specific example provides some motivation to potentially support the idea that fixing $b$ may not lead to optimal schemes in arbitrary domains, especially ones involving channels of high entropy which also portray predictable structural properties\footnote{Consider the contrived case where the last $c < b$ bits of symbol $i$ may restrict the degree of freedom of symbol $i+1$.}.

\subsubsection{Potential Extensions of the Model}

While considering the formalism of the channel in the context of provably secure steganography, 
a simple potential generalization occurred to me which I can best describe as \textit{n-out-of-k-channel steganography}.  There exist heuristic methods of making steganography harder to detect by distributing concealed messages among multiple mediums.  Using this same line of thought, would it be worthwhile to also formalize a definition of steganographic security (and also constructions) where access to $n$ out of $k$ channels is required to 
detect and retrieve concealed messages?

In parallel to the idea of proving the security of steganographic schemes, it also occurred to me that we might be able to formalize 
a technique for identifying optimality in secrecy rate for general schemes. \footnote{Perhaps this could be done by viewing the output 
of Alice as a stochastic process as in the previous section.}

\section{Exploring Appliciability of the Formalism of Provably Secure Steganography}

Having considered the current formal framework of provably secure steganography as given by 
Hopper et. al., I then moved to consider how well this framework directly admits construction of 
efficient, high-rate, and provably secure stegosystems in general domains.  This section 
contains a summary of work I encountered attempting to achieve such a construction, as well 
as a presentation of my own work to attempt to modify the formal framework to be more accommodating 
to such constructions.

\subsection{Paper: Variable-length $\mathcal{P}$-Codes}

In \cite{BadSteg}, Tri Van Le attempts to give a somewhat modified paradigm for 
provably secure steganography, and he also attempts to introduce a new steganographic 
primitive, called a $\mathcal{P}$-code, which would allow him to construct 
what he calls ``essentially optimal'' steganographic systems.  What follows in this 
section is a high-level summary of the components of this work which diverge from \cite{BiglouPSS}.
\newline\newline
Following some motivation and preliminaries, the author introduces his modified formal 
framework for provably secure steganography which is equivalent to that in \cite{BiglouPSS}, up to two 
minor modifications.  The first of these modifications is trivial, in that he introduces 
new names for the constructs already present in the framework of Hopper et. al (e.g., 
``sampler'' for oracle, the symbol $\mathcal{P}$ to represent the channel, ``chosen hiddentext 
security'' for steganographic secrecy in the presence of passive wardens).  The second of these modifications is more interesting: he models the channel as a statistical distribution over the support of a finite message space
(as opposed to fixed-length blocks of bits).  While his definition of advantage does not incorporate a 
complexity-theoretic attempt to classify wardens as in \cite{BiglouPSS}, this modification seems to be promising.

Within the bounds of this framework, Le then defines what he calles $\mathcal{P}$-codes.  A $\mathcal{P}$-code is 
a uniquiely decodable, variable-length decoding scheme $\Gamma = (\Gamma_{enc}, \Gamma_{dec})$ where the domain of 
$\Gamma_{enc}$ is $\{0,1\}^n$, the range is the message space of the channel $\mathcal{P}$, 
and the quantity

\[ \sum_{c \in \{\Gamma_e(x) \mid x \in \{0,1\}^n \}} \mid Pr_{x \sim \{0,1\}^n}[\Gamma_{enc}(x) = c] - Pr[c] ] \mid \]

\noindent is negligible in $n$.\footnote{Equivalently, that the distribution of $\Gamma_{enc}$ is indistinguishable 
from the distribution $\mathcal{P}$.}

We see through later constructions given by Le that the existence of $\mathcal{P}$-codes admits extremely 
simple steganographic schemes \footnote{The example construction by Le implements $SE$ by simply taking the exclusive-or of a hiddentext message and a random bit string $r$ and then returning the result of encoding this string with $\Gamma$.}
 \footnote{In fact, the existence of such a construction would allow ordinary cryptography to be used to arbitrarily 
 construct steganographic schemes.}.  Unfortunately, however, though Le attempts to give a valid $\mathcal{P}$-code derived from an arithmetic coding scheme, 
 there seems to be an error in his construction which could introduce vulnerability in stegosystems which use it.  In the first two 
 steps of his $\Gamma_{enc}$ procedure, he does the following:

 \begin{enumerate}
 \item{Initialize a set of variables to be used and set the intial history to be empty.}
 \item{Sample $\mathcal{C}_{h_0}$ $l$ times into $c_1,...,c_l$}
 \end{enumerate}

 \noindent This sequence of messages $c_1,...,c_l$ is then later output as the first $l$ messages of $\Gamma_{enc}$.  If Alice uses 
 $\Gamma_{enc}$ as in the example construction of Le, Ward will see $l$ messages which obey the distribution $\mathcal{C}_{h_0}$, but there is 
 no guarantee that the sequence $c_{j+1},...,c_l$ obeys $\mathcal{C}_{h_0 \mid \mid c_1 \mid \mid ... \mid \mid c_j}$ for all $j$, and so the use 
 of steganography might be easily detected.
\newline\newline
The remainder of \cite{BadSteg} concerns itself with applying $\mathcal{P}$-codes to construct public- and private-key steganographic schemes,
as well as proving that $\mathcal{P}$-codes admit ``essentially optimal'' rates in many cases. 

\subsection{An Alternative Formalism for Channels with Variable-length Messages}

Since the modified framework of provable steganographic security in \cite{BadSteg} seemed to address some concerns I found with the notion of 
universal steganography from \cite{BiglouPSS}, I thought it appropriate to attempt to reconcile the two frameworks with the goal of 
obtaining a formalism which is potentially more natural for a wider variety of domains.  This section contains the result of my attempt.
\newline\newline
\noindent \textbf{Definition of a channel } A \textbf{channel} $\mathcal{C}$ is a distribution on messages from a finite message space 
$\mathcal{M}$, parameterized by an associated set of labels $\tau$.  In other words, $\mathcal{C}$ is a statistical distribution over 
support $(c_1 \in \mathcal{M}, \tau_1),...(c_{\mid \mathcal{M} \mid} \in \mathcal{M}, \tau_n)$.  Note that this definition is a superset 
of the previous definition, as we may choose $\mathcal{M} = \{0,1\}$ and define $\tau$ such that each $\tau_i$ must contain a time label 
which is monotonically increasing among all $\tau_i, \tau_j, \tau_k$, $i < j < k$.\footnote{$\tau$ may yet contain other parameters, e.g. 
signal strength, such that said parameters may be incorporated into the definition of a stegosystem.}

Denote by $\mathcal{C}_h$ the channel distribution conditioned on a history $h$ of drawn and parameterized messages.  Further denote by 
$\mathcal{C}_h^{\Rightarrow(i)}$ the conditional distribution of the next $i$ messages in the channel given the drawn and parameterized 
history $h$.  Introduce in addition the notion of the \textit{view} of a channel: define the $q(n)$-view of the next $i$ messages 
of a channel $\mathcal{C}$ conditioned on history $h$ to be the distribution $C_h^{\Rightarrow(i)}$ with support restricted to only 
those $i$-message sequences whose length in binary representation is less than $q(n)$ bits; denote such a channel view by 
$C_h^{\Rightarrow(i)} \big\vert q(n)$.
\newline\newline
\noindent \textbf{Definition of a (symmetric-key) stegosystem }  A \textbf{stegosystem} $S$ is a pair of probabilistic algorithms
$(SE, SD)$.  $SE$ takes as input a key $K \in \{0,1\}^n$, a hiddentext $m \in \{0,1\}^*$, a history $h$, and uses an oracle $M$ capable of 
samping from $\mathcal{C}_h^{\Rightarrow(i)} \big\vert s(n)$ (for some $i$, $s(n)$ defined by the system) to return as output a sequence of messages 
$c_1 \mid \mid ... \mid \mid c_l$ from the support of $\mathcal{C}_h^{\Rightarrow(l)} \big\vert p(n)$ (where $l$ may vary between hiddentexts $m$).  
$SD$ takes as input a key $K$, a sequence of messages $c_1 \mid \mid ... \mid \mid c_l$, a history $h$, and an oracle $M$, and returns 
a hiddentext $m$.  Define correctness in a manner similar to the framework of \cite{BiglouPSS}.
\newline\newline
\noindent \textbf{Steganographic secrecy against passive wardens }  A passive warden $W$ is a warden which plays the following game:

\begin{enumerate}
\item{$W$ is given access to an oracle $M(h)$ which is capable of sampling from $\mathcal{C}_h^{\Rightarrow(i)} \big\vert v(n)$ given 
arbitrary histories $h$ and for some function $v(n)$.}
\item{$W$ is then given access to one of two oracles: }
\begin{enumerate}
\item{$O_0(m,h) \Leftarrow SE(K,m,h)$}
\item{or $O_1(m,h) \Leftarrow \mathcal{C}_h^{\Rightarrow(l)} \big\vert v(n)$}
\end{enumerate}
\item{$W$ then outputs a bit, representing whether or not he was given $O_0$ or $O_1$.}
\end{enumerate}

Denote by $(t,q,v(n))$-warden a warden which takes at most $t$ steps and $q$ queries having at most a $v(n)$-view of the 
channel while playing this game.  In this setting, the advantage of $W$, $Adv(W)$, is the quantity

\[ \mid Pr_{K,r}[W^{M,O_1} \Rightarrow 1] - Pr_{r}[W^{M,O_0} \Rightarrow 1] \mid \]

Say that a stegosystem $S$ is \textit{steganographically secret} if the advantage of any probabilistic polynomial-time warden 
is negligible in $n$.
\newline\newline
This alternative framework seems to offer three primary benefits over the previous: (1) it is clearly more natural for 
modeling channels in domains involving variable-length messages, (2) it offers the capacity to incorporate elements 
other than time into stegosystems, and (3) it partially decouples the number of bits viewed by a warden from the number of 
messages produced by $SE$.\footnote{In practice, such a coupling could result in the proof derived within a framework 
not being sound in that the system proved secure actually has a vulnerability.}

\section{Constructing and Implementing Provably Secure Steganography in Practice}

Though it may be possible to improve upon the formal framework given in \cite{BiglouPSS}, the fact 
remains that, in reality, the framework it provides is applicable to many domains.  As such, 
I spent time exploring current applications of provably secure steganography in practice.

There are potentially many motivating reasons to pursue the creation of secure steganographic 
systems; consider as a motivating scenario the world of Alice, Bob, and Ward, where Ward has a 
large degree of control of Alice and Bob.  What if Ward is, say, a nation-state government who has outlawed 
encryption?  What if Ward institutes a strict key disclosure law, where all keys used by Alice and Bob to communicate 
must be disclosed upon request?  Steganography immediately becomes relevant in these situations as a means for 
Alice and Bob to communicate confidentially.

Of course, in such a situation, the utility of steganographic systems is linked to the channels 
over which they operate.  In a scenario akin to the ones given, Ward will have a vested interest 
in disallowing steganography: even if a system exhibits a provable 0\% detection rate, Ward may 
simply outlaw the medium used.  For this reason, we would reasonably conclude that a 
steganographic system operating over an outdated landline phone protocol would be significantly 
less useful than a system which uses the English language as a medium; more generally, one might say that the
value of a steganographic system directly correlates with the indispensability 
of the underlying channel.

With this idea in mind, I began to consider what vital technologies might 
also serve as suitable steganographic mediums.  In beginning this search, I noted two things: (1) that 
computer networking is ubiquitous and fundamental to the function of the modern world, and (2) that 
\textit{secure} computer networking is ubiquitous and fundamental to the function of the modern world.
While exploring existing work attempting to use networking protocols as steganographic channels, I found 
that nearly no work attempts to provide constructions within any framework of provably secure steganography.
In considering (2), I realized that there is one common aspect to virtually every secure networking protocol
which also lends itself very well towards the formal steganographic model of \cite{BiglouPSS}: cryptographic primitives 
and cryptographic protocols \footnote{specifically the uniform randomness used by cryptographic primitives and protocols}.

This section summarizes existing work I encountered while considering the use of networking protocols 
as a steganographic medium and presents my own work in using randomness in cryptography to design and implement 
practical steganographic schemes.

\subsection{Paper: HICCUPS, Network Steganography at the Link Layer}

HICCUPS is a system presented by Szczypiorski \cite{HICCUPS} which attempts to use the data link layer to implement a 
steganographic system.  This system is included in this report as an example of a typical steganographic system in practice which 
provides no rigorous guarantee of secrecy: the entire premise of the system is the use of intentionally incorrect values in 
integrity fields (e.g. checksum fields), a technique whose use may be trivially detected by simply verifying the correctness of 
included checksums.

There may be some use in the method in highly unreliable networks; however, as the transmission probability decreases,
so does the plausibility of faking transmission errors, but, in the case of this system, there is yet no proof that the distribution 
of errors induced in the checksum will match the distribution of natural errors induced within the network.  Further, the use of this 
method arbitrarily in a network may lead to `hubs' of slightly increased transmission error probability centered around the nodes employing it, 
providing yet another possibility for detection which has not been addressed.

\subsection{Paper: Murdoch and Lewis, Network Steganography at the Internet/Transport Layer}

Diverging somewhat from the methods of the previously discussed paper, in ``Embedding Covert Channels in TCP/IP'' \cite{MurdochLewis}, Steven Murdoch and 
Stephen Lewis present an overview of common approaches to achieving steganography in the TCP/IP networking layer, concluding that most 
existing approaches are vulnerable in practice.  While their approach does not make use of the formal framework of provable steganography, they raise multiple concerns which relate closely.
\newline\newline
The authors begin with an overview of current approaches in TCP/IP steganography, stating that most existing proposals 
result in output distributions different from the distribution expected from ordinary TCP/IP.  As is the case for steganography 
at the link layer, TCP/IP steganography proposals attempt to hide information in the header fields of packets as they are constructed.
The authors discuss the fields that are most commonly used, listed below:
\newline\newline
\noindent \textbf{IP Header Fields }  IP is a common Internet-layer protocol responsible for addressing and host-to-host routing of packets.  The following headers are commonly incorporated into IP steganography proposals:
\begin{enumerate}
\item{\textit{Type of Service } The ToS field is an 8-bit field which is commonly unused.  Most platforms will, by default, 
set the ToS field to zero; thus, any use will be easily detected. }
\item{\textit{IP Identification (IPID) } The IPID is a 16-bit identifier for datagrams used during the fragmentation process.  
Though meant to be unpredictable, the IPID field is not random on most platforms, instead being a number generated by a predictable deterministic process 
depending on the host platform. There exist steganographic proposals for this field which replace the IPID with pseudo-random numbers; 
however, the authors indicate that such a method is easily detected, as the IPID is not random.}
\item{\textit{IP Flags} The (two) flags used in IP packets have a well-defined meaning; steganographic use would, aside from 
potentially inducing undefined network behavior, be easily detected.}
\item{\textit{IP Fragment Offset } When packets are fragmented, this field is used to reconstruct the original packet.  The authors 
indicate that there exist steganographic proposals which convey information by modulating packet sizes; however, this is detectable 
within many networks where such fragmentation is unusual.}
\end{enumerate}

\noindent \textbf{TCP Header Fields } TCP is a common transport-layer protocol responsible for connection-oriented, reliable host-to-host communication 
on top of the IP protocol.  The following headers are commonly incorporated into TCP steganography proposals:
\begin{enumerate}
\item{\textit{Sequence Number } The sequence number of a TCP packet is a 32-bit number which numbers the packets sent in a TCP 
connection.  Since the relationship between subsequent sequence numbers is determined, there only exists a degree of freedom in the 
choihce of the \textit{Initial Sequence Number (ISN)}.  These numbers are chosen carefully to enforce uniqueness and overlap 
constraints; however, existing proposals either replace the ISN with the hiddentext or encrypt the hiddentext (using e.g. DES) and 
include it as the ISN.  In practice, however, such a choice of ISN does not follow the platform-dependent ISN selection method and 
is therefore easily detected.}
\item{\textit{Timestamp } The timestamp field is a header field (technically two 32-bit header fields) used by some platforms (in some situations) to 
measure round-trip latency.  Use of this field for steganographic purposes is detectable in many cases by nature of the fact that many platforms 
no longer use the field; additionally, use of the field in existing proposals differs from the expected distribution.}
\end{enumerate}

In the remainder of the paper, the authors explore in detail the platform-dependent mechanisms which determine the previously 
discussed fields.  Further, they develop a suite of exhaustive tests which are used to detect uses of these fields for steganographic purposes 
(noting the previous proposals which failed these tests).  Using these tests, the authors construct a steganographic suite 
called \textit{Lathra} which is able to avoid detection in this model on OpenBSD and Linux platforms.  
\newline\newline

This work by Murdoch and Lewis emphasizes the trend in practical work in steganography to neglect to provide rigorous 
guarantees of securitiy; even more, this work reveals the extreme difficulty in constructing useful, secure steganographic 
schemes in the face of platforms which do not always obey protocol specifications.

\subsection{Embedding Steganography within Cryptographic Primitives and Protocols}

\cite{HICCUPS} and \cite{MurdochLewis} illustrated the difficulty which exists in using 
base networking protocols as steganographic channels: the varied (yet predictable) tendencies 
of their implementations to diverge from their specifications.  As a result, the underlying 
channel distribution is less easily characterized and less amenable to the framework and constructions of \cite{BiglouPSS}, 
all the while not admitting much prospect for high-rate steganographic schemes.\footnote{These methods tend to rely on header fields, usually less than 6\% of 
total packet size and consisting primarily of heavily determined, low-entropy fields.}

Not all network functionality is as forgiving of breaches of specification; among such functionalities 
are secure networking protocols, used ubiquitously to guarantee the confidentiality and safety of 
systems, people, and governments alike.  Since their removal would cause modern businesses to halt, 
banks to fail, and potentially lives to be lost, secure networking protocols certainly seem to meet 
the threshold of indispensability.  Even more promising, these protocols also share a common denominator which must be 
implemented properly and which frequently makes use of randomness: cryptography.  

In this section, I present my work on the use of cryptographic constructions used in secure networking protocols
to create steganographic channels.  I first give and prove the security of a simple stegosystem which operates over the 
initialization vectors of ciphers operating in CBC mode; I then present a system which implements this scheme and my work 
to integrate it into OpenSSL as a real-world steganographic application; finally, I present initial work I have performed 
to design stegosystems utilizing primitives in elliptic curve cryptography as steganographic channels.

\subsubsection{Steganography among Initialization Vectors in TLS 1.2/AES-CBC}

TLS (Transport Layer Security) is one of the most pervasive modern cryptographic protocols, used 
to provide confidentiality, integrity, and source verification in addition to reliable transport.  One of the 
most well-known applications of TLS is the HTTPS protocol, used to secure standard web traffic.

The TLS protocol specification exhibits a high degree of structure, itself being composed of a set of 
sub-protocols: the handshaking protocol, the cipher agreement protocol, and the application data protocol 
\cite{tls12}.  It is in this third sub-protocol, the application data protocol, that secured data 
transfer takes place; owing also to the structure of TLS, all data transferred through the application 
data protocol must pass through a single well-defined set of functionalities, the record layer, which 
determines how confidentiality and integrity guarantees are enforced \cite{tls12}.

As one might imagine, TLS guarantees confidentiality of data at the record layer through the use of encryption.  However, 
as the length of data handled by TLS is not guaranteed to fit into a single block for a given cipher specification, 
TLS necessarily employs ciphers in a variable-length mode of operation.  Conveniently, the latest non-draft specification for 
TLS, TLS 1.2, prefers the use of AES in CBC mode, and the specification states that initialization vectors (IVs) must be 
uniformly random and unpredictable per-IV \cite{tls12} \footnote{This was at least the case until recently; 
due to timing attacks on CBC, recent proposals have proposed to drop support for CBC entirely in TLS 1.3, and most TLS 1.2 implementations 
default to GCM.}.

Given that IVs must be random, the channel distribution that would correspond to the sequence of IVs seen during a sequence of 
TLS 1.2 connections is both well-defined and well-suited to the development of a stegosystem that achieves a rate of secrecy 
greater than one bit per block.  In this section, I provide a simple stateful symmetric-key stegosystem for the IVs of TLS 1.2/AES-CBC and a proof of its 
steganographic secrecy within the framework of \cite{BiglouPSS}.
\newline\newline
\noindent \textbf{A stegosystem for TLS 1.2/AES-CBC }  Let $F(K,x)$ be a secure pseudo-random permutation $F: \{0,1\}^n \times \{0,1\}^b \Rightarrow \{0,1\}^b$.
Let $N$ be a synchronized counter shared between Alice (the holder of the secret) and Bob (the receiver of the secret), initialized to 0.
Take $b$ as both the block size for $S$ and the IV size of the cipher.  Because initialization vectors 
are chosen uniformly at random in TLS 1.2/AES-CBC, we have that the conditional channel distribution $\mathcal{C}_h^b$ is the uniform distribution over support $\{0,1\}^b$ for all histories $h$.  We define the following stegosystem $S=(SE,SD)$:

\begin{algorithm}[H]
\caption{Steganographic encoding procedure}\label{1a}
\begin{algorithmic}[1]
\Procedure{$SE$}{$K \in \{0,1\}^n, m, h$}
\State Split $m$ into r-bit blocks $m_1,...,m_k$.
\For{$i$ from 1 to $k$}
\State Increment $N$.
\State \textbf{yield} $F(K,N) \oplus m_i$
\EndFor
\EndProcedure
\end{algorithmic}
\end{algorithm}

\begin{algorithm}[H]
\caption{Steganographic decoding procedure}\label{1a}
\begin{algorithmic}[1]
\Procedure{$SD$}{$K \in \{0,1\}^n, c_1 \mid \mid ... \mid \mid c_l, h$}
\For{$i$ from 1 to $l$}
	\State Increment $N$.
	\State \textbf{yield} $F(K,N) \oplus c_i$ 
\EndFor
\EndProcedure
\end{algorithmic}
\end{algorithm}

Note in the above scheme that I use the term \textit{yield} rather than \textit{return}.  This is simply done to induce the timestamp parameter 
of the bits in any stegotext block $c$ to match the timestamp of the would-be innocent initialization vector.
\newline\newline
\noindent \textbf{Proof of Steganographic Secrecy } This proof makes use of the well-known result that the advantage of a 
secure PRP $F(K \in \{0,1\}^n, x \in X = \{0,1\}^b)$ is negligible in the PRF setting with $q$ queries if $\frac{q^2}{\mid X \mid}$
is negligible.
\newline\newline
\noindent \textit{Claim } $S$ is steganographically secret against any probabilistic polynomial-time warden when $n=b$.
\newline\newline
Say that there exists a $(t(n), q(n), l(n))$-warden $W$ which achieves non-negligible advantage in the game for 
steganographic security given in \cite{BiglouPSS} where $t(n)$, $q(n)$, and $l(n)$ are polynomials in $n$.  In other words, that the quantity 

\[ Adv_{S,\mathcal{C}}(W) = \mid Pr_{K,r}[W_r^{M,SE(K,m,h)} = 1] - Pr_{r}[W_r^{M,O(m,h)} = 1] \mid \]

is non-negligible for this warden.  In the steganographic secrecy setting, consider a party $A$ which simulates either $SE(K,m,h)$ or $O(m,h)$ 
by playing the PRF security game as follows:

\begin{enumerate}
\item{The broker flips a single coin $r$.  If $r=1$, the broker chooses a key $K \in \{0,1\}^n$ for $F$.  Else, 
the broker chooses a truly random function $F'$.}
\item{$A$ initializes a counter $N$ to 1.}
\item{$A$ then waits for a queries from $W$.  On query $m$, $A$ queries the broker with $N$ and obtains $v = F(K,N)$ or $v = F'(N)$.  
$A$ then responds to $W$ with $v \oplus m$ and then increments $N$.}
\item{Once $W$ outputs a guess, $A$ forwards this as its guess to the broker.}
\end{enumerate}

Note that, in the third step, if $r=1$, the response to $A$ is precisely the value of $SE(K,m,h)$; if $r=0$, the response to $A$ is random, 
since $v = F'(N)$ is random, and therefore a simulation of $O(m,h)$ given our characterization of $\mathcal{C}_h$.  Since $A$ makes as many 
queries as $W$ and as much time as $W$ to output a guess, we know that $A$ is polynomial both in time taken and queries made.

Consider the quantity $\mid Pr[A \Rightarrow 1 \mid r = 1] - Pr[A \Rightarrow 1 \mid r = 0] \mid$; this is the advantage of $A$ in the 
PRF security game against $F$.  The value $Pr[A \Rightarrow 1 \mid r = 1]$ is precisely the probability that the warden outputs 
1 given that $SE(K,m,h)$ is being simulated; the value $Pr[A \Rightarrow 1 \mid r = 0]$ is precisely the value that $O(m,h)$ is 
being simulated.  Therefore, the advantage of $A$ in the PRF security game is exactly the advantage of $W$ in the setting for 
steganographic secrecy; thus, if $W$ has non-negligible advantage against $S=(SE,SD)$, then $A$ has non-negligible advantage against 
$F$ in the PRF security game.  This directly contradicts our choice of $F$: by the previously stated result, we know that 
the advantage of any probabilistic polynomial-time adversary is $< \frac{q(n)^2}{2^{n}}$ when $b=n$, therefore negligible.
\newline\newline
\noindent Taking $n = b = 128$, the IV size for AES, we conclude that $S$ is steganographically secret against 
all probabilistic polynomial-time wardens in our application (and that therefore the probability of detection is less than 
$\frac{poly(128)}{2^{128}}$) \footnote{In practice, then, we would probably want to switch keys after many uses of $SE$.}.
 
\paragraph{Steg-MQ, a Steganographic Message Queue}

Having designed and proved a stegosystem suitable for use in TLS 1.2, I next endeavored to integrate it into a real-world application.
I completed this task in two steps: (1) implementation of a \textit{steganographic message queue} and (2) modification of the OpenSSL
implementation of TLS (version 1.1.1) to integrate the change.

Towards (1), I implemented a system called \textit{Steg-MQ}.  Steg-MQ is a steganographic message queue: in other words, it's a process
which runs on a machine and provides the following functionality to other processes:

\begin{enumerate}
\item{The ability to \textit{publish} data to be hidden according to a channel distribution and steganographic scheme.  Published 
stegotexts are stored into one of multiple stegotext queues, each designated for use by a specific application which handles the transmission 
of the stegotext.}
\item{The ability to \textit{consume} and retrieve channel messages that have been published into an application's stegotext queue (ideally to then 
be sent by the application).}
\item{The ability to \textit{decode} incoming stegotexts and publish them into one of multiple application-specific hiddentext queues.}
\item{The ability to \textit{retrieve} and consume decoded hiddenteexts from an application's hiddentext queue.}
\end{enumerate}

\noindent Systems like Steg-MQ are necessary for the practical of implementation in practice.  While stegosystems may be sound in 
theoretical formulation, real-time transmission of the output of $SE$ requires available cover (e.g., a packet being sent 
via TLS).  Another practical area of concern is duplication of implementation; by centralizing the implementation of steganographic 
operations to a single service, we can reduce inevitable error that would occur from multiple applications independently implementing them.

The current rudimentary form of Steg-MQ (implementing only \textit{publish} and \textit{consume} for a single global queue and 
for only the channel distribution and stegosystem defined in this section) may be found on GitHub (https://github.com/ad-alston/steg-mq).
\newline\newline
After implementing the proposed stegosystem within Steg-MQ, and after implementing a simple dynamic library that may be used by 
any application to interface with Steg-MQ, integration required only a minimal change to the OpenSSL implementation of TLS (which may also 
be found on GitHub: https://github.com/ad-alston/openssl-steganography).  The interaction between Steg-MQ and OpenSSL's implementation of 
TLS 1.2 is depicted in the following sequence diagram:

\begin{figure}[H]
\centering
\caption{Sequence Diagram for Steg-MQ and OpenSSL}
\includegraphics{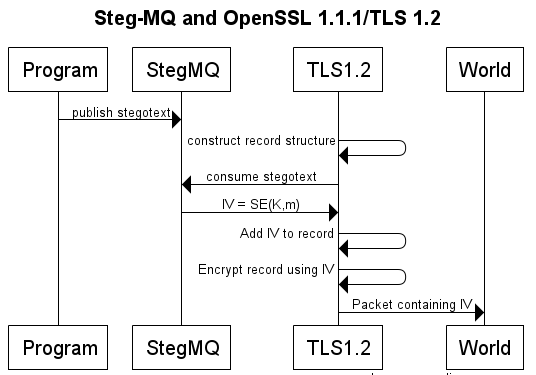}
\end{figure}

\subsubsection{Steganography among Elliptic Curves}

After designing and implementing a steganographic system that operates over channels of initialization vectors, 
I began to explore the use of other cryptographic primitives as steganographic channels.  Most recently, I have been 
exploring the possible use of ephemeral-key elliptic curve Diffie-Hellman exchanges and the elliptic curve digital 
signature algorithm to this end.  This section presents a high-level overview of the approaches I have been considering.

\paragraph{Ephemeral-key Elliptic Curve Diffie-Hellman and Elliptic Curve Digital Signature Algorithm}

Say that Alice and Bob are preparing to perform an ephemeral-key elliptic curve Diffie-Hellman exchange for a curve 
whose domain parameters include a generator $G$ of order $n$.  I present a stateful stegosystem $S = (SE,SD)$ which operates on a 
channel distribution $\mathcal{C}$ over ephemeral keys generated within this domain.  Let 
$F(K,x)$ be a PRP $\{0,1\}^k \times \{0,1\}^n \Rightarrow \{0,1\}^n$ and $H_K(K,x)$ be a PRP $\{0,1\}^k \times \{0,1\}^r \Rightarrow \{0,1\}^r$.  
Let $N$ be a synchronized counter set to 0.

\begin{algorithm}[H]
\caption{Steganographic encoding procedure for ECDH}\label{1a}
\begin{algorithmic}[1]
\Procedure{$SE$}{$K \in \{0,1\}^n, m, h$}
\State Split $m$ into r-bit blocks $m_1,...,m_k$.
\For{$i$ from 1 to $k$}
\State Set $d = d^{(H(K,m_i))}$, where $d^{(j)}=F(K,N+j)$.
\State \textbf{yield} ($d,dG$)
\State Increment $N$.
\EndFor
\EndProcedure
\end{algorithmic}
\end{algorithm}

\begin{algorithm}[H]
\caption{Steganographic decoding procedure for ECDH}\label{1a}
\begin{algorithmic}[1]
\Procedure{$SD$}{$K \in \{0,1\}^n, c, h$}
\State Enumerate $d^{(j)}=F(K,N+j)$ until $d^{(j)}G = c$.
\State If no such $j$ found, quit.
\State \textbf{return} $H^{-1}(K,j)$. 
\State Increment $N$.
\EndProcedure
\end{algorithmic}
\end{algorithm}

Note that the above procedure allows the transfer of $r$ bits of transformation per elliptic curve communicated.  In the context 
of ECDH, the scheme would be applied as follows:

\begin{enumerate}
\item{Alice, wanting to communicate an $r$-bit message, computes her ephemeral key as $SE(K,m) = (d_A, Q_A)$.}
\item{Bob generates his ephemeral key as usual (or uses $SE$ himself with another key and counter).}
\item{Alice sends Bob her ephemeral public key  $Q_A$.}
\item{Bob runs $SD(K,Q_A)$ and obtains the message.}
\item{Alice and Bob complete ECDH as usual.}
\end{enumerate}

No direct proof of secrecy is given (for the sake of the length of this document); however, the intuition behind such a proof would be that, 
as $d_A$ should be pseudo-random, the distribution over the public component of the public key should be indistinguishable from the innocent case.

Since an explicit enumeration of $2^r$ values is required, in practice, such a scheme may only ever be implemented with 
$r$ of 10-15 (with precomputation and other clever implementation tricks).  This may still prove useful, however: consider a 
generic web service which runs on HTTPS and experiences 10-20 requests per second.  Such a web service may be able to covertly
communicate megabytes of information per day by means of ECDH alone.  Such a rate of secrecy would additionally be more than enough for 
two parties to establish a covert key for future communication in, say, 3-4 connections.

I also note that the above approach is also relevant to the elliptic curve digital signature algorithm (ECDSA) by a similar application.
\newline\newline
\textbf{A note on ECDSA } If it can be shown that the distribution of the x-coordinates of points $P = dG$ where $d$ is chosen uniformly 
at random is itself uniformly random \footnote{I admit that this may simply not be the case for some or any curves.}, one may obtain a higher 
rate of secrecy by simply obtaining random curve points as follows:

\begin{enumerate}
\item{Apply a PRP to hide message $m$ (whose length in bits is equal to that of the x-coordinate of points on this curve).}
\item{Obtain the y-coordinate by solving the proper quadratic equation.  (If there is no point for the chosen x, 
modify the scheme by using the last $k$ bits of the message as padding and increment through all $2^k$ padding values until 
the point is valid.)}
\end{enumerate}

\section{Conclusion and Next Steps}

Over the course of this project, I have explored publications relating to the semantic definition of provably secure 
steganography, and I have surveyed available work on implementing efficient provably secure steganography in practice.  As a 
result of this investigation, I myself devised and presented a natural modification to the formal framework of provably 
secure steganography, and I have designed and implemented provably secure steganographic constructions in a highly 
relevant real-world context to address the lack of such systems in practice.

The continuation of this work would rest in considering all of the points considered in greater depth, in so doing addressing one 
or all of the following ideas:
\newline\newline \noindent \textbf{The formal framework of provably secure steganography: } Is there a natural and useful domain where 
all of the frameworks given in this paper are not sound or are exceedingly cumbersome?  Is there some domain where one is and one is not?
Is there some way to formalize a measure of optimality of rate based upon a stochastic model of communication?
\newline\newline \noindent \textbf{Extensions of steganography: } Where might a construction of \textit{n-out-of-k-channel steganography} be useful?
An ambitious goal may be to give a construction of \textit{n-k steganography} in this domain.
\newline\newline \noindent \textbf{Steganography in practice: } Further substantiate the implementation of Steg-MQ.  Implement the scheme for 
elliptic curve steganography within Steg-MQ and augment OpenSSL to use it.  Use Steg-MQ to integrate one of these schemes (potentially the 
scheme for ECDSA) in another application (potentially a blockchain technology).
\newline\newline \noindent \textbf{Cryptography as a channel: } Investigate other opportunities to take advantage of randomness in cryptographic 
primitives for the sake of steganography; explore the truth behind relevant questions (such as the one regarding the distribution 
of x-coordinates: is this true in some curves?).
\newline\newline \noindent \textbf{More ambitious channels for steganography: } Find more ambitious channels that don't rely so much 
on uniform randomness.  Attempt to develop and implement a provably secure stegosystem which uses something like web technologies, mark-up 
languages, program execution patterns, or even natural language as a steganographic channel. 

\bibliographystyle{acm}
\bibliography{report}

\end{document}